\documentclass[journal]{IEEEtran}
\usepackage{graphicx}
\usepackage{epsf}
\usepackage{mathtools}
\usepackage{amssymb}
\usepackage{amsmath, amsfonts}
\usepackage{latexsym}
\usepackage{subfigure}
\usepackage{multirow}
\usepackage{multicol}
\usepackage[table]{xcolor}
\usepackage{color}

\usepackage{tabularx}
\usepackage{lipsum}

\usepackage{algorithm}
\usepackage{algorithmic}

\usepackage{tikz}
\usepackage[hidelinks]{hyperref}
\usepackage{url}

\allowdisplaybreaks

\definecolor{orange}{RGB}{255,127,0}
\definecolor{blue}{RGB}{0,0,255}
\definecolor{red}{RGB}{220,0,0}
\definecolor{green}{RGB}{0,120,0}
\definecolor{grey}{RGB}{255,120,255}
\definecolor{purple}{RGB}{100,0,140}

\begin{document}

\title{\fontsize{20}{24}\selectfont Simulations for Stochastic Geometry on the Performance of V2X Communications in Rural Macrocell Environment}

\author
{
Victor Obi and Seungmo Kim, \textit{Senior Member}, \textit{IEEE}

\thanks{V. Obi and S. Kim are with Department of Electrical and Computer Engineering at Georgia Southern University in Statesboro, GA, USA. The corresponding author is S. Kim who can be reached at seungmokim@georgiasouthern.edu.}
}

\maketitle

\begin{abstract}
Vehicle-to-everything (V2X) communications is a concept that has been around for the past decade. It involves communication between vehicles and other types of infrastructure. This application is exceptionally useful for emergency services such as ambulances, fire trucks etc. This is because an emergency vehicle can communicate with the traffic light infrastructure and make it give the green signal thereby allowing vehicle to pass quickly. This is useful because it alerts other cars and pedestrians on the road when an emergency vehicle is present.  In this paper, a V2X communications system in an urban setting will be simulated using MATLAB and the Automated Driving Toolbox. The purpose of simulation in MATLAB is to test if vehicle to vehicle communication is affected by buildings and other infrastructure. The first Simulation is constructed using the Automated driving simulator. In this simulation clover highway is constructed and cars are added to mimic a busy highway. The second simulation involves several nodes being programmed in MATLAB and intersecting at various points to simulate an overpass highway. The end goal of this project to successfully simulate a V2X communications system with additional components added within the program to represent additional cars and infrastructure.
\end{abstract}


\begin{figure*}
\centering
\includegraphics[width = \linewidth]{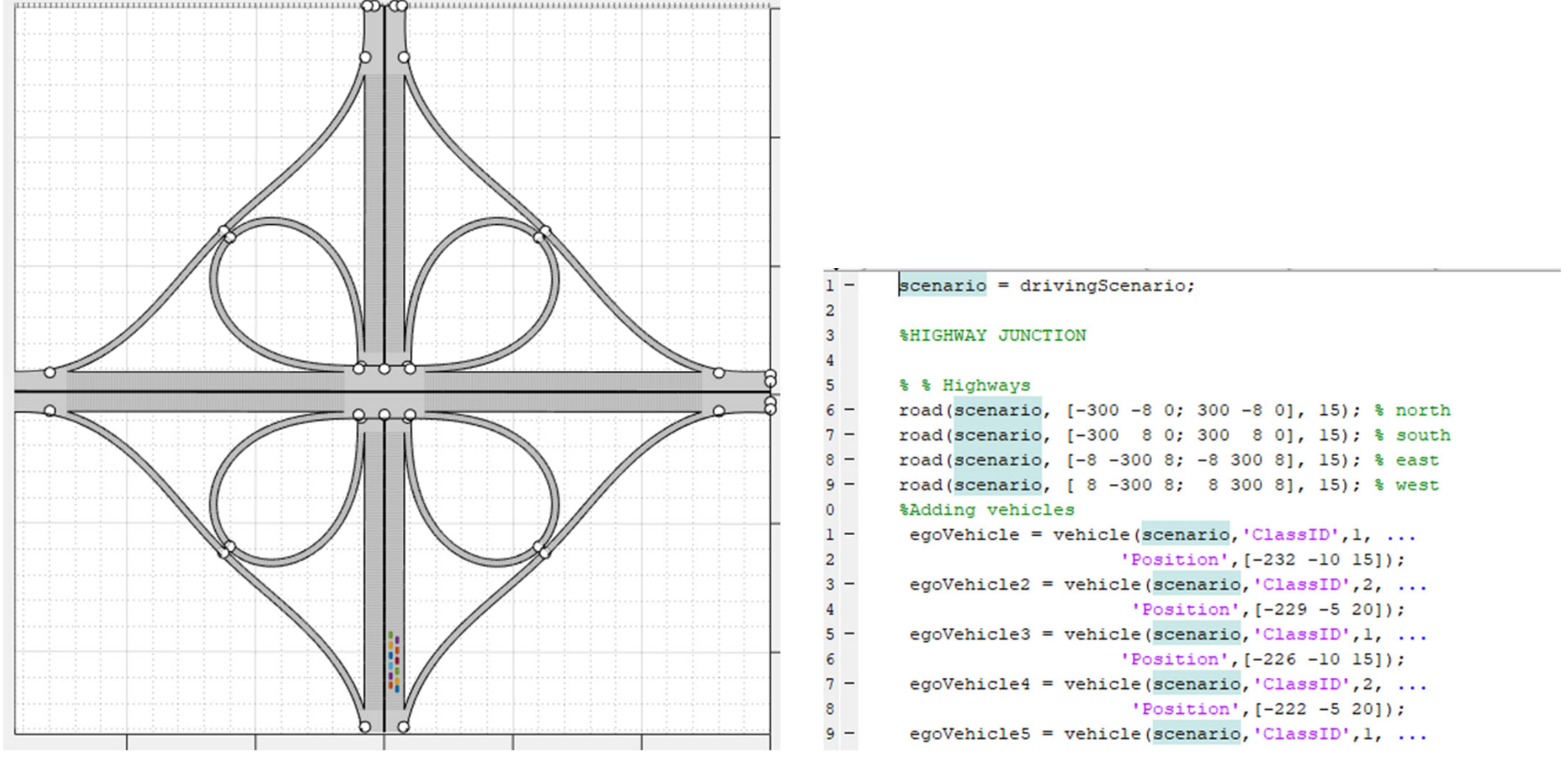}
\caption{The original approach for the project}
\label{fig_original}
\end{figure*}

\section{Introduction}\label{sec_intro}
The purpose of this paper is to construct a clover highway with vehicles on all roads and have movement by the cars on all junctions. The project will be accomplished using MATLAB and the toolboxes associated with it. Through this project we will simulate vehicle-to-everything communications (V2X) communications which stands for vehicle to everything communications. There are several components to V2X, which include Vehicle to vehicle (V2V), vehicle to infrastructure (V2I), vehicle to pedestrian (V2P), vehicle to network (V2N), etc. As such, V2X has already taken the notion as a leading driver technology enabling cars to communicate with each other and other external systems (i.e. traffic lights, streetlights, parking garages, pedestrians) \cite{access19}.

Nevertheless, the V2X communications are at the most critical moment in its history. The main reason is the federal government's movement that favors the Wi-Fi as an effort to meet the burgeoning bandwidth demand. The movement directly affected the spectrum for V2X communications. The spectrum of 5.850-5.925 GHz (also known as the \textit{5.9 GHz band}) was assigned to V2X communications for supporting the intelligent transportation system (ITS) use cases in 1999 by the United States Federal Communications Commission (U.S. FCC). However, after many years of hearing and discussion, the FCC voted in a bipartisan support to reduce the bandwidth for V2X from 75 MHz (5.850-5.925 GHz) to 30 MHz (5.895-5.925 GHz) while granting the remainder of the 5.9 GHz to Wi-Fi \cite{nprm}.

In particular, a V2X communications system has been found to be interference-constrained mainly due to a large communications range for each vehicle \cite{etri17}. There were two main technologies supporting the V2X communications functionalities, namely the dedicated short-range communication (DSRC) \cite{globecom18}\cite{access20} and C-V2X \cite{vtc20}\cite{vtc21}. Moreover, such an inter-technology interference between dissmilar wireless systems has been studied quite a few times in the litereature \cite{pimrc08}\cite{lett17}. Albeit not presented in this paper, our analysis has found that a suburban geometry yielded a higher level of interference because of the \textit{openness} in signal exchange among vehicles. On the other hand, ironically, an urban geometry marked a lower level of interference among vehicles due to existence of \textit{blockage} provided by physical obstacles (e.g., buildings). Motivated from this finding, this paper focuses on the performance evaluation of a V2X communications system in a suburban environment.

\textit{Contribution of This Paper:} This paper provides a suburban highway environment where the performance of a V2X communications can be evaluated. The framework aims to serve as a conceptual basis for further study based on the stochastic geometry, which will more precisely assess the performance of V2X communications in safety-critical use cases. As a means to accommodate a higher level of flexibility in the distribution of vehicles, this study is focused on providing a computer simulations framework.

\begin{figure*}
\centering
\includegraphics[width = \linewidth]{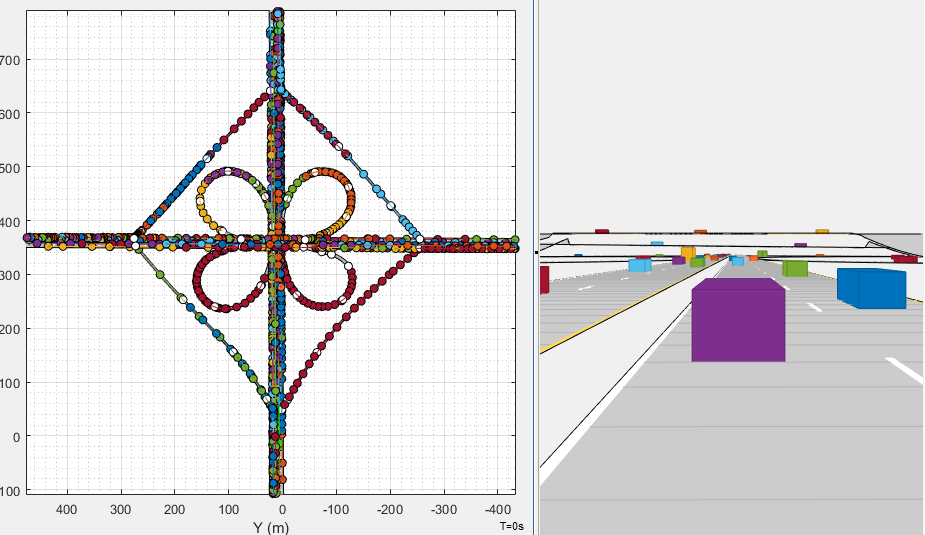}
\caption{Highway Simulation using automated driving toolbox (The blocks represent the vehicles and the dots represent the paths the vehicles take.)}
\label{fig_highway}
\end{figure*}

\begin{figure*}
\centering
\includegraphics[width = \linewidth]{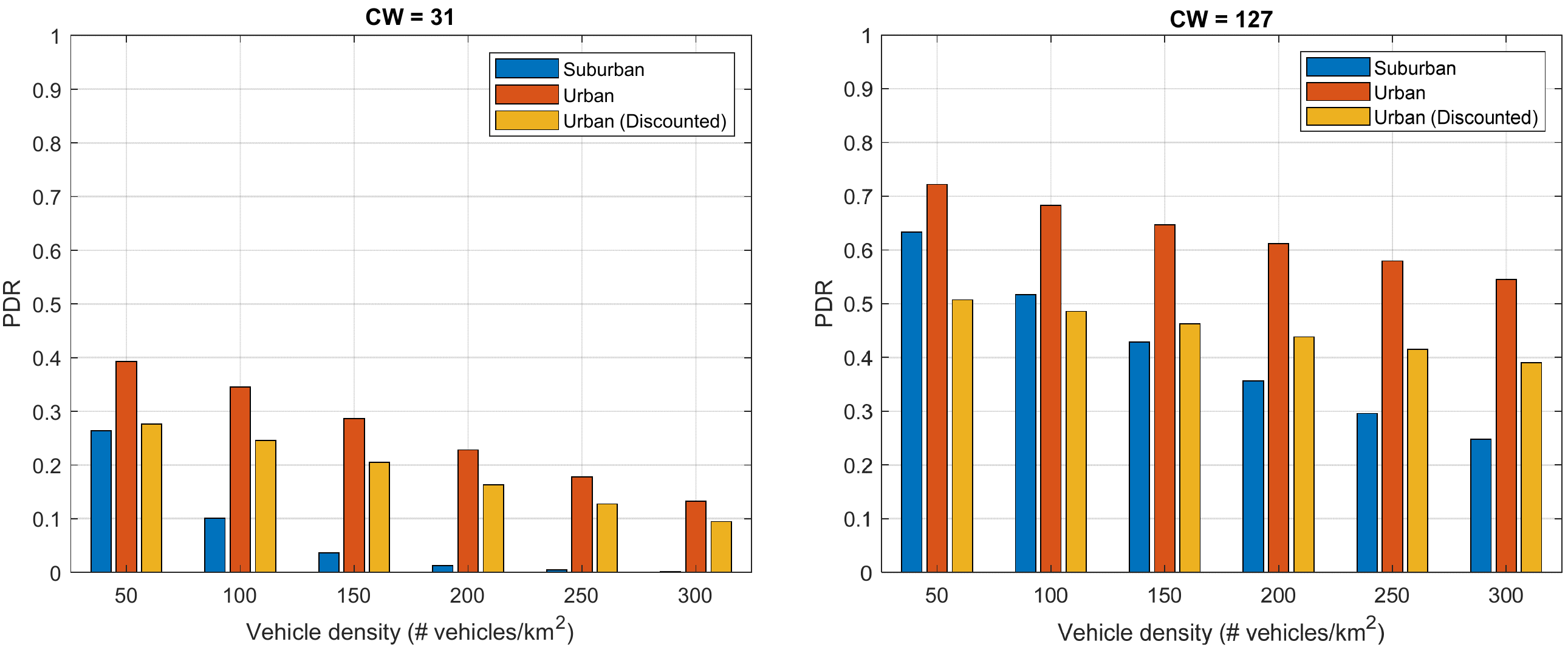}
\caption{Comparison of PDR between urban and suburban geometries (with CW = \{31, 127\})}
\label{fig_result}
\end{figure*}

\section{System Model}\label{sec_model}
Recall from Section \ref{sec_intro} that a suburban environment is of particular significance in the analysis of V2X communications due to a higher interference. As an effort to capture the unique characteristics of a suburban geometry, this paper adopts a \textit{Rural Macrocell (RMa)} environment that is defined in the channel model by the 3rd Generation Partnership Project (3GPP) \cite{tr38901}. Technical details follow in this section.

\subsection{Path Loss}
The rural deployment scenario focuses on larger and continuous coverage. The key characteristics of this scenario are
continuous wide area coverage supporting high speed vehicles. Several key technical characteristics are provided in the specification \cite{tr38901}:
\begin{itemize}
\item \textit{Distribution of nodes:} Uniform
\item \textit{Indoor/Outdoor:} 50\% indoor and 50\% outdoor
\item \textit{LOS/NLOS:} Both LOS/NLOS
\end{itemize}
where LOS and NLOS stand for line-of-sight and non-line-of-sight, respectively.

For the path loss model characterized for the RMa geometry, the readers are encouraged to refer to Table 7.4.1-1 of the specification \cite{tr38901}.

\subsection{Geometry}
Now, we proceed to creation of geometric models for the RMa. As a representative of the RMa model, we create a suburban environment where two highway segments cross, which forms a 4-way junction. Fig. \ref{fig_original} describes the geometry (on the left) and a snapshot of the MATLAB code implementing the geometry (on the right).

On the created geometry, we move on to distribute vehicles at the density of 100 vehicles per km$^{-2}$, which is illustrated in Fig. \ref{fig_highway}. The normalization by km$^{2}$ is attributed to the \textit{dimension} of the geometry as shown on the righthand side of Fig. \ref{fig_highway}. Details on our methodology will follow in the next section.

\section{Simulations Methodology}
This section depicts \textit{why} and \textit{what} we do in the simulations as proposed in the previous section.

\subsection{Justification of Methodology}
We found that simulation would accomplish the best efficiency as the main method to evaluate the performance of the proposed mechanism, based on several advantages \cite{ibrahim20}.

First, as presented through Figs. \ref{fig_original} through \ref{fig_highway}, the parameters defining and operating the proposed study are quite diverse in types and values, which makes it challenging to explore the parameters' dynamic orchestration in concert. A simulation provides a relatively easier control over such a large space composed of various parameters with wide ranges of values. It gives an obvious advantage over mathematical derivations and testbed implementations. One can easily anticipate a high degree of complexity in changing the setting every time a new round of simulation is executed, while a large number of iterations is inevitable to present a statistically stable result in such a complex setting. As an effort to deal with the complexity, we adopt simulation as the main methodology, which, as shall be presented in this section, did efficiently evaluate the proposed system in a wide diversity of parameter settings.

Second, simulations enable computations without being caught up with restrictions or errors caused by computing environmental factors including hardware, compiler, language, etc. Taking into account all the available options for all of those factors will complex the performance evaluation to a too high degree, which, as such, will make it hard to precisely identify the factors determining the key performance.

\subsection{Development of Simulations Framework}
During the early stages of the project, there was an attempt to construct the highways in MATLAB programmatically however that proved to be in efficient for the goal in mind (as shown in Fig. \ref{fig_original}). The approach for the project was then changed and instead of creating the roads programmatically we decided to draw it out using the Automated driving toolbox (as seen in Fig. \ref{fig_highway}). Through use of the automated driving toolbox, the highways were constructed, and the moving vehicles were able to be simulated.

\section{Results}
Taking a comparative look at the urban and suburban geometries, we calculate the performance of a V2X network in terms of the packet delivery rate (PDR) as shown in Fig. \ref{fig_result}. The figure shows that with CW of 31 and 127, a vehicle can achieve a higher PDR in an urban scenario. The rationale is that that the blockage acts to reduce the number of neighboring vehicles that ``compete'' for a channel. In other words, in an urban scenario, the receivability among the Rx vehicles not undergoing blockage is higher compared to a suburban setting. The reason is that the blockage also serves as physically dividing a large network into smaller ones.

However, we would not recommend one taking the result in Fig. \ref{fig_result} that an urban setting is more advantageous in the performance of exchanging signals. This is because the PDR is not displaying the number of vehicles that received the BSM. In other words, the physical \textit{coverage} of a message broadcast must be suppressed in an urban setting compared to a suburban scenario where no blockage exists. Therefore, the third bar (``orange'' in color) in the figure shows the PDR that is ``discounted'' by the blockage rate, which gives the number of vehicles that are actually able to receive a BSM after consideration of the blockage.

\section{Conclusion and Future Work}
This paper has presented a simulation framework that provides an analytical capability based on the stochastic geometry. The paper has put particular focus on the suburban geometry, in consideration of a higher interference caused by lack of blockage of signals among vehicles. Based on the simulations, this paper has also presented a comparative analysis between the urban and suburban scenarios. The result indicated that a higher PDR could be achieved in an urban setting, but a discount must be applied because the higher performance was achieved only among a certain subset of vehicles due to the blockage.

Moving forward the plan is to add more nodes at various points of the graph and vary the speeds for each node to see how that affects the performance of the simulation. Though certain aspects of the code have been manipulated to help reach the goal of consistency, the plan is to push the boundaries of the code and see what other applications the code can be used in.


\end{document}